# GAMMA RAY LINES FROM THE ORION COMPLEX[1]

REUVEN RAMATY

Laboratory for High Energy Astrophysics

Goddard Space Flight Center

Greenbelt, MD 20771

BENZION KOZLOVSKY

School of Physics and Astronomy

Tel Aviv University

Tel Aviv, Israel

RICHARD E. LINGENFELTER

Center for Astrophysics and Space Sciences

University of California San Diego

La Jolla, CA 92093

## ABSTRACT

We show that the 4.44 and 6.13 MeV line emission observed with COMPTEL from Orion is consistent with gamma ray spectra consisting of a mixture of narrow and broad lines or spectra containing only broad lines. We employed several accelerated particle compositions and showed that the current COMPTEL data in the 3–7 MeV region alone cannot distinguish between the various possibilities. However, the COMPTEL upper limits in the 1–3 MeV band favor a composition similar to that of the winds of Wolf-Rayet stars of spectral type WC. The power dissipated by the accelerated particles at Orion is about $4 \times 10^{38}$ erg s$^{-1}$. These particles are not expected to produce significant amounts of $^{26}$Al.

---

[1] In press, The Astrophys. J. Letters

1. INTRODUCTION

Gamma ray line emission in the 3 to 7 MeV range was observed from the Orion complex with COMPTEL on the Compton Gamma Ray Observatory (Bloemen et al. 1994). The radiation shows emission peaks near 4.4 and 6.1 MeV, consistent with the deexcitation of excited states in $^{12}$C and $^{16}$O. These lines can only be produced by accelerated particle interactions (Ramaty, Kozlovsky and Lingenfelter 1979). Even though gamma ray lines produced by accelerated particles have been seen from many solar flares (e.g. Chupp 1990), this is the first instance that such lines are observed from a galactic source. The Orion spectrum, however, is very different from solar flare spectra. Whereas Orion reveals only upper limits below 3 MeV, solar flare spectra show strong emission in this region, both in lines (mainly due to Ne, Mg, Si, S and Fe) and electron bremsstrahlung (Murphy et al. 1991).

Bloemen et al. (1994) suggested that the excitations are produced by the accelerated C and O nuclei themselves, rather than by accelerated protons and $\alpha$ particles bombarding ambient C and O. Bykov and Bloemen (1994) proposed that the interactions are predominantly due to nuclei around 30 MeV/nucleon accelerated by colliding stellar winds and supernova explosions, and suggest that abundances resulting from supernova explosions could explain the COMPTEL upper limits below 3 MeV. Nath & Biermann (1994) also pointed out that the stellar winds of the young O and B stars in Orion will produce shocks which are efficient particle accelerators. Clayton (1994) suggested that the same particles which produce the gamma rays in Orion will lead to an enhanced $^{26}$Al production which could be sufficient to account for both the total galactic $^{26}$Al mass implied by gamma ray observations (e.g. Diehl et al. 1993) and the $^{26}$Mg excess seen in meteorites in the early solar system (Lee, Papanastassiou & Wasserburg 1977).

In this letter we present results obtained by employing a detailed nuclear deexcitation code to analyze the COMPTEL Orion observations. This code, originally developed for application to gamma ray line production in the interstellar medium (Ramaty et al. 1979), has been extended and used to analyze solar flare gamma ray emission (Murphy et al. 1991). The code incorporates a large number of nuclear reactions, and it allows us to calculate deexcitation line spectra in either a thin or a thick target model, for various ambient and accelerated particle compositions, and various accelerated particle energy spectra. We address the following questions: do the 3–7 MeV data really favor an accelerated particle population at Orion that is enriched in C and O relative to protons and $\alpha$ particles; what are the implications of the absence of observed emission below 3 MeV on the particle composition and energy spectrum; what information do the shapes of the lines provide on the spectrum of the accelerated particles; what other gamma ray lines could be observed from Orion; what is the total power in accelerated particles; and is the $^{26}$Al production in Orion significant? We also comment on the differences between the solar flare and Orion gamma ray spectra which might yield clues on the nature of the acceleration mechanism.

## 2. ANALYSIS

To calculate the gamma ray line production we must make assumptions on both the properties of the accelerated particles and the interaction model. The accelerated particle spectra and compositions predicted by various acceleration theories depend on many parameters, especially in the low energy range ($\lesssim$ 100 MeV/nucleon). Since the observed gamma ray lines from Orion are predominantly due to particles in this energy range, for simplicity we have chosen a spectral form which characterizes the source spectra $N_i(E)$ of the various accelerated particle species in terms of a single parameter. We take $N_i(E) = \text{const}$ for $E < E_c$ and $\propto E^{-10}$ at higher energies. While other spectral forms are also possible, we show that by varying $E_c$ over a broad range (2–100 MeV/nucleon), we are able to fully explore the dependence of main features of the gamma ray emission (i.e. its spectrum and normalization) on the hardness of the accelerated particle spectrum.

The other important property of the accelerated particles is their composition which depends on not only the acceleration mechanism but also the ambient medium from which the particles are accelerated. Rather than relying on the predictions of acceleration mechanisms, we take the ratios of the $N_i$'s at the same energy per nucleon proportional to any one of the four compositions shown in columns 3 through 6 of Table 1. We assume that the composition of the ambient medium is that of the solar photosphere given in column 2 of this table. The cosmic ray source composition (CRS) is from Mewaldt (1983) except $^1$H for which we took the solar value relative to $^4$He. The SN15M$_\odot$ and SN25M$_\odot$ compositions represent the ejecta of the 15M$_\odot$ and 25M$_\odot$ supernovae given by Weaver & Woosley (1993). The WC composition represents the extreme late phase winds of Wolf-Rayet (WR) stars of spectral type WC. In such stars He, C and O are found in great abundance but there is no evidence for H (Abbott & Conti 1987). The WC values given in Table 1 for $^4$He, $^{12}$C, $^{16}$O, $^{20}$Ne, $^{22}$Ne and $^{26}$Mg are based on calculations (Maeder & Meynet 1987, figure 13) for a 60M$_\odot$ star. The other WC abundances given in Table 1 are solar values normalized to $^{20}$Ne, except $^1$H for which we arbitrarily set the abundance equal to that of $^{16}$O. In this WC composition, accelerated $^1$H plays essentially no role in gamma ray production.

We evaluate the gamma ray line production in a thick target model in which the accelerated particles produce gamma rays as they are slowed down by Coulomb energy losses in a neutral medium until their energies fall below the thresholds of the various nuclear reactions. This model is energetically more efficient than a thin target model in which the particles escape from the target region after having lost only a small fraction of their energy. Furthermore, because the rate of energy loss increases with increasing nuclear charge, the contribution of the accelerated Ne–Fe nuclei, which produce gamma ray lines mostly below 3 MeV, is suppressed in a thick target relative to the contributions of C and O which produce lines mainly above 3 MeV.

Bloemen et al. (1993) have presented an upper limit on the ratio of the 1–3 MeV to 3–7 MeV emission, $R = 2 \int_{1\text{MeV}}^{3\text{MeV}} E_\gamma^2 Q(E_\gamma) dE_\gamma / \int_{3\text{MeV}}^{7\text{MeV}} E_\gamma^2 Q(E_\gamma) dE_\gamma < 0.13(2\sigma)$, where $Q(E_\gamma)$ is the differential gamma ray emissivity in Orion. We have calculated $R$ for the four accelerated particle compositions and the results are shown in Fig. 1 (panel a) together with the COMPTEL upper limit. We see that the WC composition yields very low $R$'s which are consistent with the data for practically any accelerated particle spectrum. This is because the 1–3 MeV band is populated mostly by nuclear lines from Ne–Fe, and these nuclei are essentially absent in the WC composition. The other compositions yield higher $R$'s. The discrepancy between the data and both the CRS and SN15M$_\odot$ values is greater than $3\sigma$ at all $E_c$'s, and increases rapidly below 5 MeV/nucl. The SN25M$_\odot$ values are also inconsistent with the data, except for $E_c$'s near 5 MeV/nucl.

In panel (b) of Fig. 1 we show the ratio of the gamma ray line yield to the total energy in accelerated particles for the four compositions. Here $W = \Sigma_i A_i \int E N_i(E) dE$, where $A_i$ is atomic number. We see that the efficiency increases with increasing $E_c$ and that it is the highest for the WC composition, which is more efficient, by a factor of about 7, than the other compositions.

In Fig. 2 we compare the COMPTEL data with calculated spectra obtained by convolving calculated high resolution spectra with a Gaussian ($\sigma = 0.125$ MeV) that approximates the energy resolution of the COMPTEL instrument. Both the CRS and SN15M$_\odot$ spectra (panel a) provide acceptable fits to the 3–7 MeV data, although, as shown in Fig. 1, these spectra overproduce in the 1–3 MeV band. In panel (b), the SN25M$_\odot$ spectrum is probably unacceptable because it does not produce enough emission around 4.4 MeV. On the other hand, the WC spectrum in panel (b) fits the data quite well, and, as shown in Fig. 1, this spectrum is also consistent with the 1–3 MeV upper limit.

In Fig. 3 we show our calculated high resolution spectra for the CRS and WC compositions normalized to the observed flux in the 3–7 MeV range (Bloemen et al. 1993). The CRS spectrum shows a variety of narrow lines superposed on broad lines. The fact that this spectrum fits the 3–7 MeV data quite well (Fig. 2 panel a) shows that it is not necessary to suppress the narrow line component as suggested by Bloemen et al. (1993). The strongest predicted narrow line fluxes for the CRS composition are given in Table 2, where the indicated energy ranges contain the predicted photon fluxes. The WC spectrum (Fig. 3 panel b) consists mostly of broad lines, particularly those of $^{12}$C and $^{16}$O. The strongest narrow line is at 1.275 MeV from $^{22}$Ne whose predicted flux is about $10^{-6}$ photons cm$^{-2}$ s$^{-1}$. The confirmation of the absence of narrow Ne–Fe lines would provide support for our suggestion that the observed broad C and O lines from Orion indeed originate in WR stars.

We have also calculated the 0.511 MeV line flux resulting from positron production by the accelerated particle interactions (Kozlovsky et al. 1987). Taking the positronium fraction equal to 0.9 (Guessoum, Ramaty & Lingenfelter 1991) we obtain fluxes of $2 \times 10^{-5}$ and $1.5 \times 10^{-5}$ photons cm$^{-2}$ s$^{-1}$ for the CRS and WC compositions, respectively. The 0.511 MeV line and its associated positronium continuum could probe the physical conditions of the annihilation site (Guessoum et al. 1991).

The widths of the broad lines depend on $E_c$. We found that for the WC composition $E_c$ should not exceed about 20 MeV/nucleon because otherwise the broad 4.4 MeV line will no longer fit the data. Such an upper limit cannot be placed on the spectra which also include narrow lines (e.g. the CRS spectrum). However, if $E_c$ exceeds about 50 MeV/nucleon then spallation reactions will produce a feature around 5.2 MeV (from $^{15}$N and $^{15}$O) for which there is no evidence in the data.

In Fig. 4 we show the ratio of the $^{26}$Al yield to the 3–7 MeV photon yield based on cross sections given by Kozlovsky et al. (1987). Clayton (1994) estimated an $^{26}$Al production in Orion of about $5 \times 10^{51}$ nuclei per $10^6$ years. Using the results of Fig. 4 for $E_c >$5MeV/nucl and the observed 3–7 MeV flux of $10^{-4}$ photons cm$^{-2}$ s$^{-1}$, we obtain production rates which are lower by factors of 10 to 300, depending on the composition. It is thus unlikely that the accelerated particles in Orion play a major in $^{26}$Al production.

### 3. DISCUSSION AND SUMMARY

We have shown that the observed Orion data in the 3–7 MeV range can be fit by gamma ray spectra which consist of narrow and broad lines or by spectra which contain only broad lines. A typical example of the former is the spectrum produced by accelerated particles with the cosmic ray source (CRS) composition. On the other hand, to produce a gamma ray spectrum consisting of only broad lines, it is necessary to suppress the protons and $\alpha$ particles relative to the accelerated heavier nuclei. Such a suppression has never been encountered in any of the well studied accelerated particle populations, e.g. solar energetic particles (SEP) or cosmic rays. Nor is such a suppression predicted by known acceleration theories. As we have suggested, a possible scenario is the acceleration of the C and O rich winds of WR stars by shocks. However, none of the known WR stars (e.g. Lang 1991) is in Orion. But it is possible that some WR stars are hidden in or behind clouds or were active there in the near past.

We found that the compositions of the ejecta of 15M$_\odot$ and 25M$_\odot$ supernovae do not offer any advantages. The former is in conflict with the 1–3 MeV upper limit at about the same level as the CRS composition, while the latter, because of its low C abundance, does not produce a good fit to the 3–7 MeV data (see Fig. 2, panel b).

For the WC composition with $E_c = 20$ MeV/nucleon, the 3–7 MeV gamma ray yield is about 6 photons per erg (Fig. 1 panel b). Within the present constraints this yield is close to the maximum. The observed 3–7 MeV flux of $10^{-4}$ photons cm$^{-2}$ s$^{-1}$ then implies that the power dissipated by the accelerated particles at Orion is about $4 \times 10^{38}$ erg s$^{-1}$. In the WC scenario, for a mass loss of $5 \times 10^{-5}$ M$_\odot$ per year and a wind velocity of 4000 km s$^{-1}$ (near the maximal values given by Abbott & Conti 1987), the available kinetic energy is about $2.5 \times 10^{38}$ erg s$^{-1}$ per star. If the conversion efficiency of wind kinetic energy into shocks and accelerated particles is close to unity, a few WR stars could account for the observed gamma ray line emission in steady state. But such stars are presently not observed. Alternatively, if the wind particles are accelerated over the entire WC phase of $\sim 10^5$ yrs and a significant fraction of them subsequently diffuse into contact with dense clouds where their stopping time is much less than their acceleration time, then a single recent WR star could have produced all of the accelerated particles responsible for the emission and no longer be visible.

We pointed out that the solar flare gamma ray spectra show much higher ratios of 1–3 MeV to 3–7 MeV fluxes than does Orion. It was shown (Murphy et al. 1991) that this enhanced emission below 3 MeV is, in part, due to the enrichment of the flare accelerated particle population in heavy nuclei. Such enrichments are routinely seen in direct observations of SEP from impulsive flares (e.g. Reames, Meyer & von Rosenvinge 1994). These impulsive SEP events are also rich in relativistic electrons. On the other hand, in gradual SEP events the composition is coronal and the electron-to-proton ratio is low. The acceleration in impulsive events is thought to be due to gyroresonant interactions with plasma turbulence while in gradual events it is the result of shock acceleration. The fact that the ratio of bremsstrahlung-to-nuclear line emission in Orion is very low lends support to the shock acceleration scenario.

We wish to acknowledge the role of R. J. Murphy in the development of the nuclear deexcitation line code and thank A.G.W. Cameron for reminding us of the contribution of $^{23}$Na to $^{26}$Al production. We also wish to acknowledge financial support from the US-Israel Binational Foundation and NASA under Grant NAGW 1970.

TABLE 1  Abundances (Numbers of Nuclei Relative to $^{16}$O)

|  | Solar | CRS | SN15M$_\odot$ | SN25M$_\odot$ | WC |
|---|---|---|---|---|---|
| $^1$H | 1.18x10$^3$ | 2.19x10$^2$ | 1.45x10$^2$ | 4.38x10$^1$ | 1 |
| $^4$He | 1.18x10$^2$ | 2.19x10$^1$ | 2.71x10$^1$ | 9.87 | 9.2x10$^{-1}$ |
| $^{12}$C | 4.27x10$^{-1}$ | 8.86x10$^{-1}$ | 2.77x10$^{-1}$ | 1.10x10$^{-1}$ | 6.9x10$^{-1}$ |
| $^{14}$N | 1.32x10$^{-1}$ | 3.82x10$^{-2}$ | 7.39x10$^{-2}$ | 2.62x10$^{-2}$ | 2.4x10$^{-3}$ |
| $^{16}$O | 1.00 | 1.00 | 1.00 | 1.00 | 1.00 |
| $^{20}$Ne | 1.28x10$^{-1}$ | 8.30x10$^{-2}$ | 1.46x10$^{-1}$ | 2.01x10$^{-2}$ | 2.3x10$^{-3}$ |
| $^{22}$Ne | 1.56x10$^{-2}$ | 3.54x10$^{-2}$ | 7.97x10$^{-3}$ | 1.02x10$^{-2}$ | 2.0x10$^{-2}$ |
| $^{23}$Na | 2.51x10$^{-3}$ | 1.71x10$^{-2}$ | 3.94x10$^{-3}$ | 5.21x10$^{-4}$ | 0 |
| $^{24}$Mg | 3.53x10$^{-2}$ | 1.46x10$^{-1}$ | 2.23x10$^{-2}$ | 5.69x10$^{-3}$ | 6.3x10$^{-4}$ |
| $^{26}$Mg | 5.01x10$^{-3}$ | 3.32x10$^{-2}$ | 5.94x10$^{-3}$ | 2.97x10$^{-3}$ | 4.0x10$^{-3}$ |
| $^{27}$Al | 3.74x10$^{-3}$ | 2.70x10$^{-2}$ | 3.91x10$^{-3}$ | 1.23x10$^{-3}$ | 6.7x10$^{-5}$ |
| $^{28}$Si | 3.83x10$^{-2}$ | 1.66x10$^{-1}$ | 7.97x10$^{-2}$ | 7.38x10$^{-2}$ | 6.9x10$^{-4}$ |
| $^{32}$S | 1.81x10$^{-2}$ | 2.50x10$^{-2}$ | 4.23x10$^{-2}$ | 3.15x10$^{-2}$ | 3.3x10$^{-4}$ |
| $^{40}$Ca | 2.70x10$^{-3}$ | 1.30x10$^{-2}$ | 5.22x10$^{-3}$ | 2.34x10$^{-3}$ | 5.0x10$^{-5}$ |
| $^{56}$Fe | 5.04x10$^{-2}$ | 1.61x10$^{-1}$ | 4.08x10$^{-2}$ | 4.26x10$^{-2}$ | 9.1x10$^{-4}$ |

TABLE 2. Predicted Narrow Line Fluxes for the CRS Composition Normalized to the Observed 3–7 MeV Flux

| Energy Range (MeV) | Isotope | Photons cm$^{-2}$ s$^{-1}$ |
|---|---|---|
| 0.40–0.50 | $^7$Li–$^7$Be | $1.8 \times 10^{-5}$ |
| 0.475–0.48 | $^7$Be | $3.1 \times 10^{-6}$ |
| 0.84–0.854 | $^{56}$Fe | $5.5 \times 10^{-6}$ |
| 1.232–1.244 | $^{56}$Fe | $3.4 \times 10^{-6}$ |
| 1.61–1.66 | $^{20}$Ne | $6.7 \times 10^{-6}$ |
| 4.35–4.52 | $^{12}$C | $1.6 \times 10^{-5}$ |
| 6.0–6.25 | $^{16}$O | $1.2 \times 10^{-5}$ |

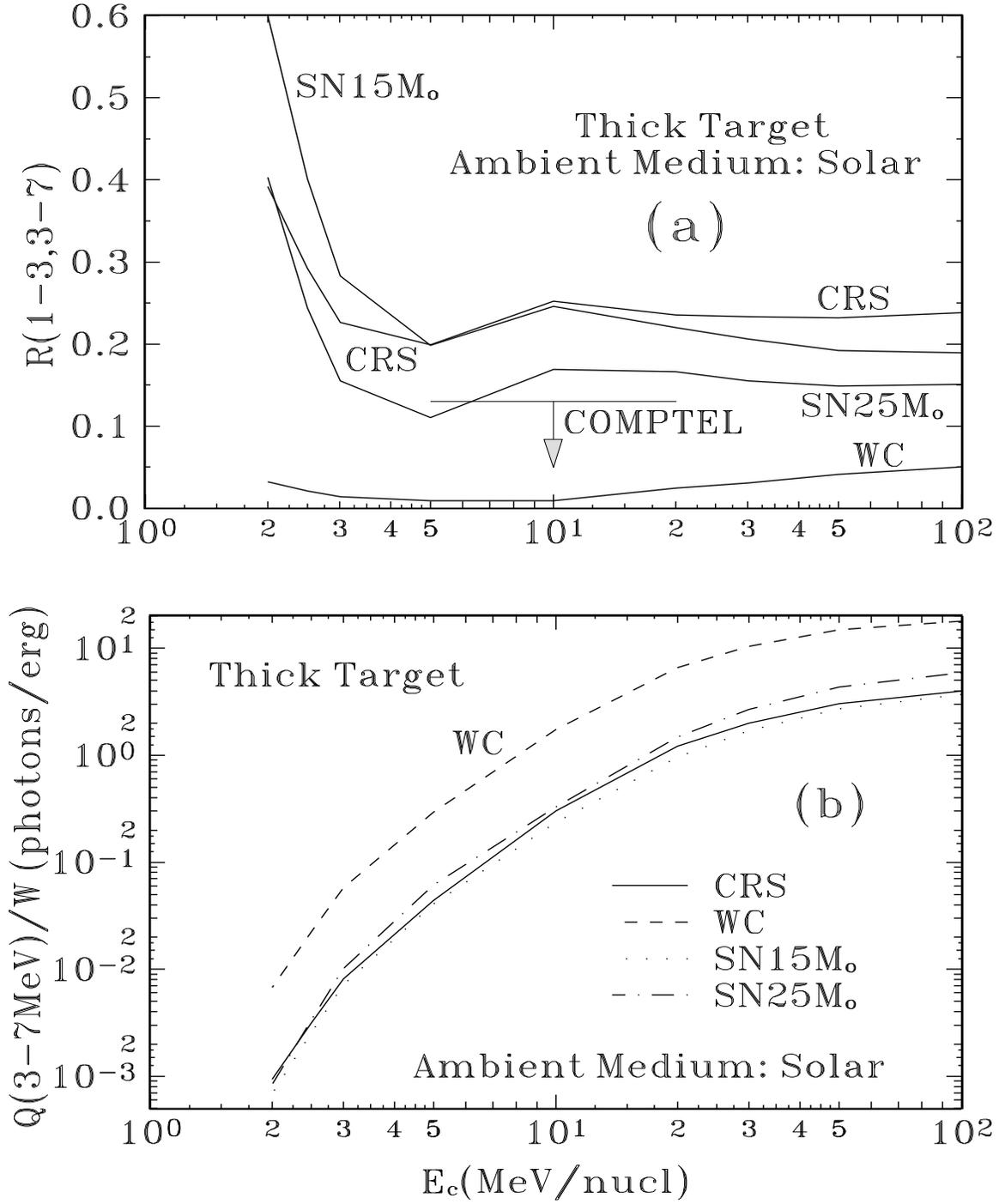

Fig. 1. (a): Ratio of 1–3 to 3–7 MeV emission for various compositions; $R$ and $E_c$ are defined in the text. (b) The 3–7 MeV gamma ray yield per erg deposited for the various compositions.

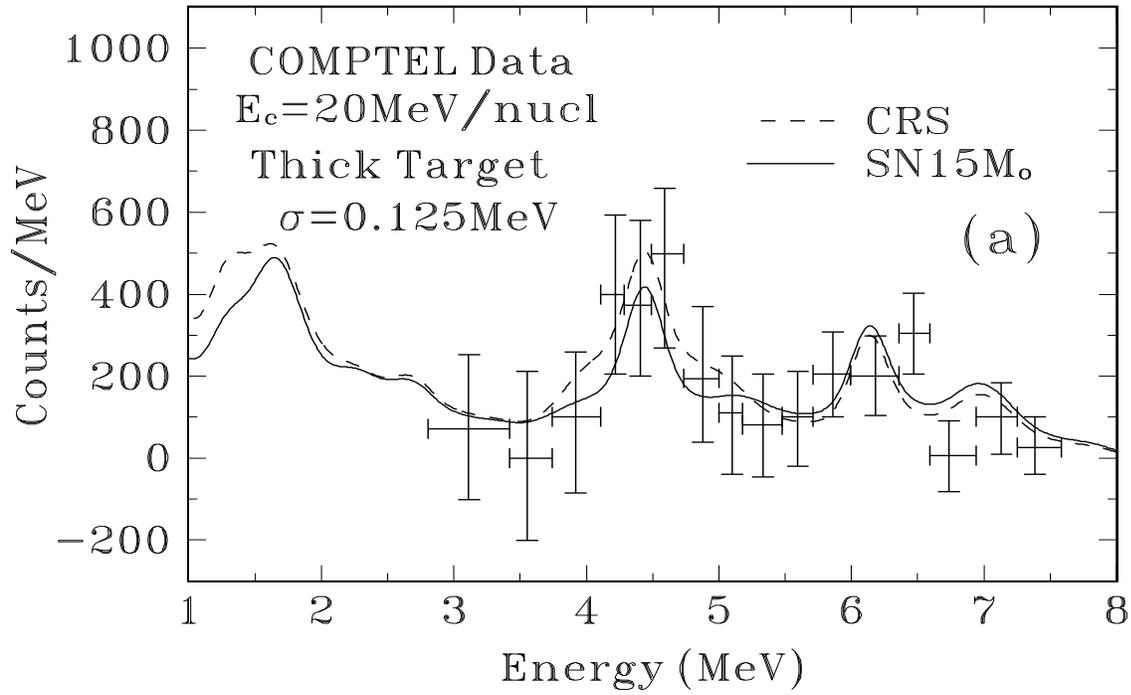
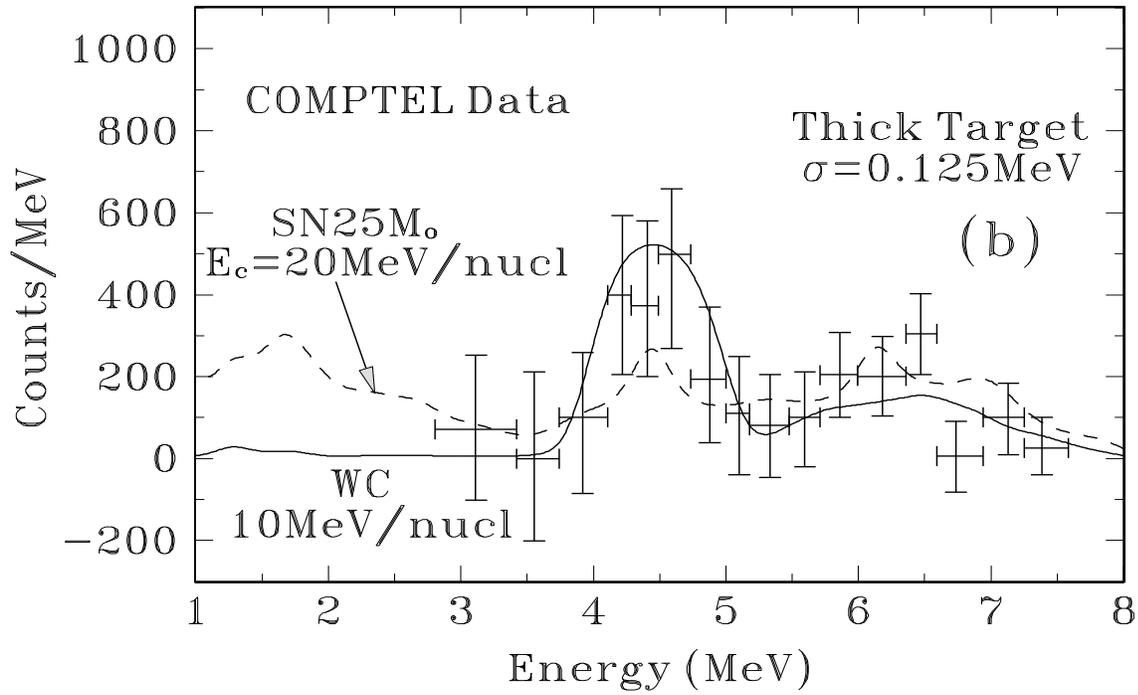

Fig. 2. Fits to the COMPTEL data for the various compositions.

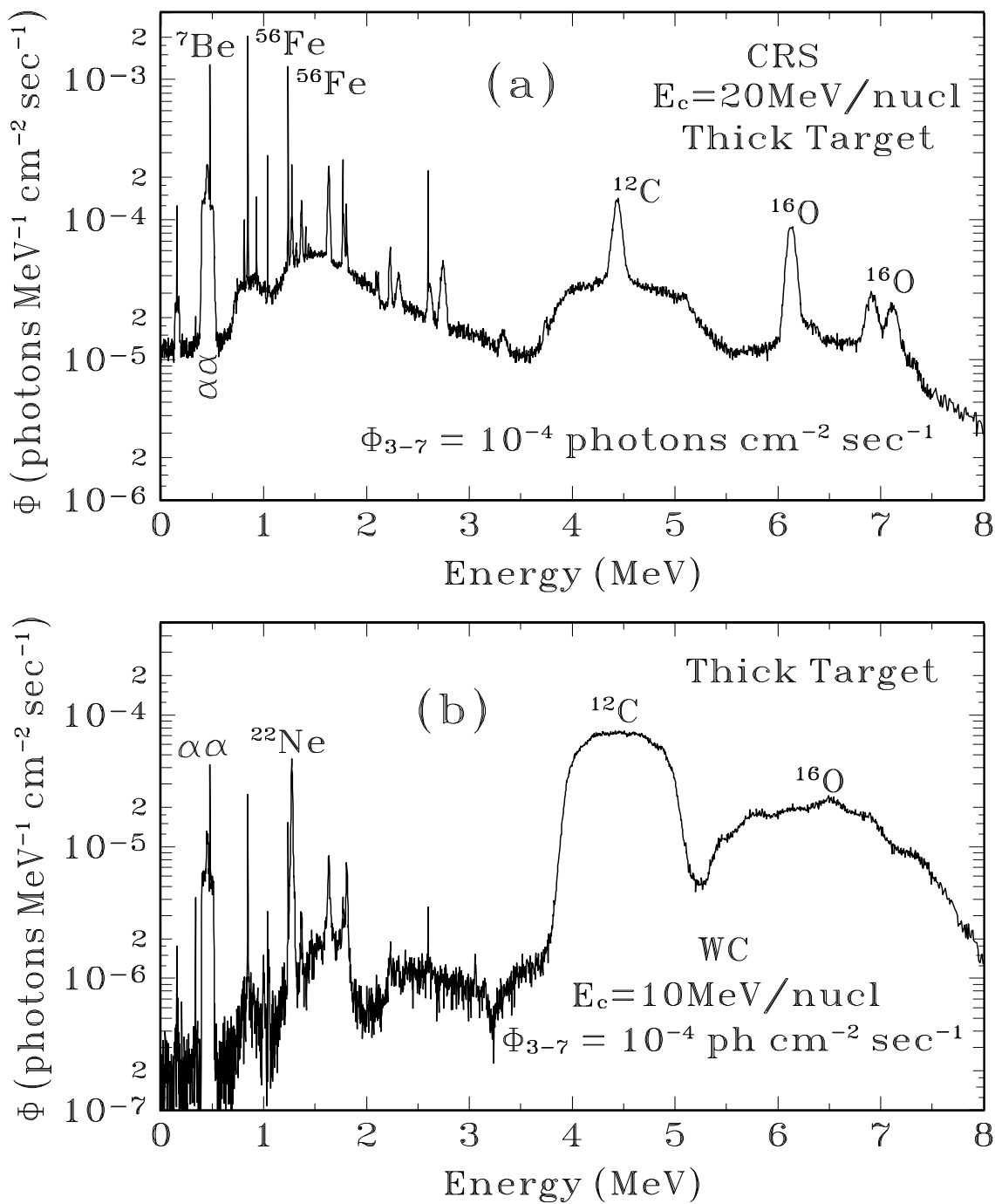

Fig. 3. Calculated high resolution spectra. The comparison of the WC (Wolf-Rayet) and CRS (cosmic ray source) cases shows the effects of the suppression of protons and Ne–Fe relative to C and O in the accelerated particle population.

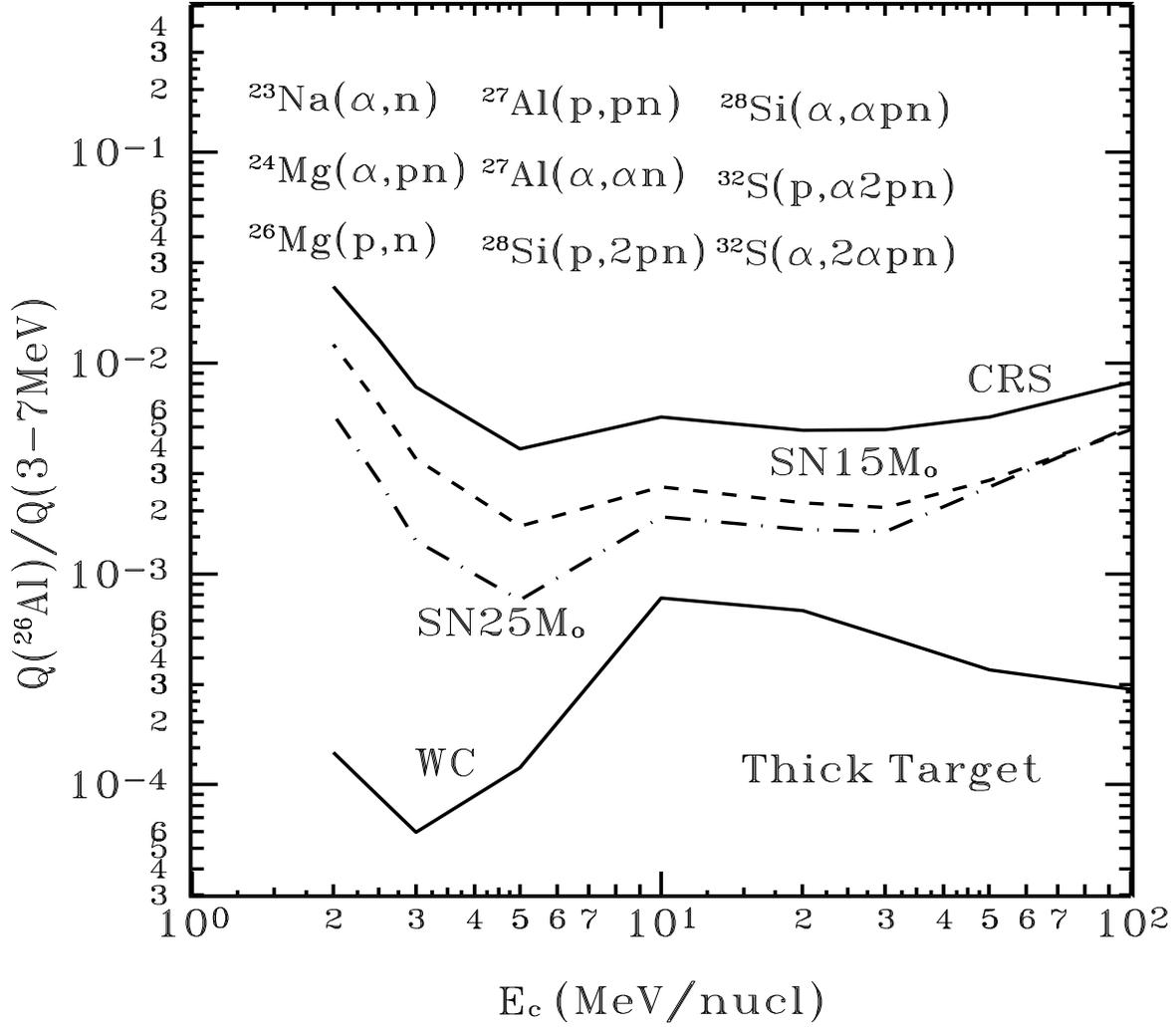

Fig. 4. The ratio of $^{26}$Al production to 3–7 MeV photon production for the various compositions. The $^{26}$Al producing reactions are listed in the figure.